# An Incentivized Approach for Fair Participation in Wireless Ad hoc Networks

Arka Rai Choudhuri
Department of Computer Science and Engineering
National Institute of Technology, Karnataka (NITK)
Surathkal, India
ark.10co16@nitk.edu.in

Kalyanasundaram S
Department of Computer Science and Engineering
National Institute of Technology, Karnataka (NITK)
Surathkal, India
kal.10co107@nitk.edu.in

Shriyak Sridhar
Department of Computer Science and Engineering
National Institute of Technology, Karnataka (NITK)
Surathkal, India
shr.10co82@nitk.edu.in

Annappa B
Department of Computer Science and Engineering
National Institute of Technology, Karnataka (NITK)
Surathkal, India
annappa@ieee.org

*Abstract*— **In Wireless Ad hoc networks (WANETs), nodes separated by considerable distance communicate with each other by relaying their messages through other nodes. However, it might not be in the best interests of a node to forward the message of another node due to power constraints. In addition, all nodes being rational, some nodes may be selfish, i.e. they might not relay data from other nodes so as to increase their lifetime. In this paper, we present a fair and incentivized approach for participation in Ad hoc networks. Given the power required for each transmission, we are able to determine the power saving contributed by each intermediate hop. We propose the Fair and Incentivized Ad hoc Protocol (FASTER), which takes a selected route from a routing protocol as input, to calculate the worth of each node using the cooperative game theory concept of 'Shapley Value' applied on the power saved by each node. This value can be used for allocation of Virtual Currency to the nodes, which can be used for subsequent message transmissions.**

*Keywords*— **Cooperative Game theory, Fair and Incentivized, Shapley Value, Virtual Currency**

## I. INTRODUCTION

Wireless Ad hoc Networks (WANETs) are the future of mobile communication. Being infrastructure-less networks, WANETs will be able to provide the flexibility and robustness required to completely support Ubiquitous Connectivity.

An Ad hoc network is one that consists of mobile nodes that communicate over a wireless medium. Due to limited radio transmission range, data is usually forwarded by multiple intermediate hops to the destination. This action also helps in conserving the energy of the sender node, and is vital for the functioning of an Ad hoc network. The major constraint in such networks is the battery life of the nodes in the system. Since each of the nodes is battery powered, they must conserve their available power so as to avoid premature discontinuation of their activities, as energy cannot be replenished in a short time. This limitation can lead to the 'selfish' behavior of nodes, where a node refuses to forward the data of some other node in the network in order to conserve its own battery life.

Initially, Ad hoc networks were used by armies near the battlefield to carry out swift communications without any preinstalled infrastructure. In this scenario, the nodes did not need any incentive to relay another node's request, as all nodes in the network had a common purpose. Modern day Ad hoc networks are mostly formed by nodes that do not collectively try to serve a predefined purpose; their aim is to send their messages while consuming as little energy as possible. Selfish nodes are common in Ad hoc networks as each node is autonomous and wants to maximize its yield from the network.

In this paper, we argue that a node forwarding packets for some other node must gain some tangible benefit from the activity. This would act as an incentive to transmit the message to other nodes, thus helping the network thrive.

This paper describes an incentivized game-theoretical approach to develop a fair algorithm when routing information from one node to another is given. The idea that nodes providing a service must be rewarded while those accepting a service must be charged has been explored before [1, 2, 3]. Based on this theme, many acceptance algorithms have been proposed. At each node, the decision of whether to forward or reject a relay request is taken by the acceptance algorithm.

In this paper, we assume that network nodes have their own application needs and physical constraints, and thus have a choice of becoming selfish. On this basis, we aim to study the impact of selfish users on the performance of the network. It must be noted that selfish users are different from malicious users, whose aim is to disrupt the functioning of the group, even if the costs of achieving its goal are high.

To provide a fair mechanism, where the remuneration that each node receives is proportional to its contribution and importance in the network, we propose the FAir Share incenTivizEd Ad hoc paRticipation protocol (FASTER) that uses the cooperative game theory concept of Shapley Value to allocate a payoff for each service provider. The payoff to each node is determined by the amount that each user contributes to the group. The contribution of each user is calculated by the amount of aggregate transmission power that the user saves by agreeing to relay the message. The power reduction can be used as an indicator of the value of a node in the network. Thus, if a node saves a greater amount of power by forwarding data, then it will earn a greater payoff. The concept of Shapley Value was used to calculate the payoff that each node which serves the network deserves.

The payoff that each node earns will be in the form of some Virtual Currency [4]. This currency can be used as a payment for sending packets to other nodes on the network. A node that does not possess enough currency will be unable to send its own data over the network, and will thus have to be very cooperative in order to earn suitable wealth for transmission of its own information.

The proposed protocol will not take care of the route selected, and will run over a suitable routing protocol such as AODV [5] which will use appropriate routing metrics [6] to calculate the best path for transmission. The protocol can be divided into 2 parts: Utility Measurement, which will measure the contribution of each node to the network and use the concept of Shapley Value to decide the payoff of each network, and Virtual Currency Allocation, which will allocate Virtual Currency tokens to each node that provides its services.

The remainder of this paper is organized as follows: In Section 2, we review related work on the topic of cooperation in Ad hoc networks; Section 3 presents the design of FASTER; Section 4 describes the Simulation of FASTER and the results obtained; and Section 5 concludes the work done in the paper and points out areas for future research.

## II. RELATED WORK

CAP-SV [7] uses a similar mechanism of Shapley Value for rewarding nodes appropriately in a network, but also takes charge of finding efficient routes, and doesn't concern itself with currency allocation. CAP-SV proves to be an improvement on AODV in many cases.

PARO [8] uses the fact that intermediate relays help greatly in conserving the energy of the sender. Any node running PARO must always keep its radios on, and this could lead to a very significant operation cost. To counter this, protocols such as SPAN [9] have been designed. [10] describes a protocol to ensure energy efficiency in a network with selfish users by introducing the 'sympathy' parameter for forwarding traffic.

Early research works in the field of WANETs took the cooperation between different nodes for granted[1]. This primitive model is not applicable today as most Ad hoc networks consist of autonomous independent users. Hence, methods have been proposed to encourage cooperation, which greatly improves the usage that each node can get out of the network. [11] also provides a mechanism whereby currency and services are shared by different hops till the message reaches the destination.

Game Theory approaches for finding solutions to common issues in WANETs are becoming more popular today. It gives us an opportunity to treat the series of transactions between any two nodes as a game, and judging from the strategy of our opponent, allows us to make decisions. [12] proposes a game theory based protocol called GTFT that allows a node to stay in the network only if it cooperates. By this protocol, the throughputs of selfish users fall drastically.

## III. SYSTEM DESIGN

FASTER is designed to allow a fair approach to routing in Ad hoc networks. We assume that the received signal strength can be measured by each user, and that the link between any 2 nodes is bidirectional. The FASTER header would require the addition of power and previous hop information so as to enable the necessary calculations and function correctly. We also assume that all nodes on the network are completely autonomous and can choose their next action. They can also decide to be non-truthful. Finally, we presume that all wireless communications that take place between the nodes follow the two-ray ground reflection model, which is an acceptable assumption in the given conditions [13].

Every Ad hoc network can be modeled as a coalition of players, where each node in the system is a player. The aim of any coalition is to ensure that, as a group, the members receive more benefits than they would have if they had participated alone. Similarly, in Ad hoc networks, nodes will be able to transmit more messages if they cooperate with each other and forward messages. That being said, nodes would need remuneration for offering their services to the group. The payoff can be in the form of Virtual Currency, which nodes can use for transmitting their own messages.

To ensure that the mechanism is fair, the compensation that each node receives must be proportional to its contribution to the network. To achieve this, we introduce the cooperative game theory concept of Shapley Value, which assigns a unique distribution of the total reward generated by a coalition of players. As some members contribute more than others, it is only fair that the payment they receive must be greater than that received by others. Thus, we can say that each node is rewarded in accordance with its importance in the arrangement. Shapley Value is also the only formula that satisfies the three axioms of cooperative games: efficiency, symmetry, and additivity. The Shapley Value function can be calculated as shown below:

---

[1] This is primarily because early WANETs were deployed in emergency situations such as war and flood, and all nodes had a common cause

$$\phi_i(v) = \sum_{i \in S} \frac{(|S|-1)!(n-|S|)![v(S)-v(S-\{i\})]}{n!} \quad (1)$$

In the equation, 'n' is the total number of nodes in the coalition, and |S| is the current size of the coalition. 'v(S)' is the minimum reward that each member of the coalition can expect. Summation is taken over all considered members of the group and their worth is calculated. The total gain is distributed, and if two players are equivalent, i.e. v(S U {i}) = v(S U {j}), then their payoffs will be equal.

*A. Power Saving Through Forwarding*

By the two-ray ground reflection model, the power required to transmit a message via many hops is less than that required while transmitting directly from source to destination. This is because the received power is inversely proportional to distance from the transmitter by the proportionality:

$$P_r \alpha \frac{1}{d^4} \quad (2)$$

As a result of this proportionality, we can see that a change in the distance will greatly affect the received power, all other factors such as antenna heights and gains remaining constant. This reduction in transmission power can have a profound effect on network performance by increasing the lifetime of the system. By FASTER design, the sender of any message will have an idea of the amount of energy saved by each of the nodes that offer their services.

*B. FASTER Framework*

FASTER is a protocol that ensures fair participation in Wireless Ad hoc networks by using cooperative game theory approaches. FASTER does not take care of route discovery, evaluation and choosing, rather it works over a common routing protocol such as AODV which takes into account appropriate routing metrics for wireless situations before selecting a route. Given a predefined route selected by a routing protocol, FASTER will provide an incentivized mechanism where each participant feels adequately remunerated for the services offered. FASTER is composed of two stages: Partnership Formation (PF) and Virtual Currency Allocation (VCA).

*C. Partnership Formation*

The Partnership Formation phase of FASTER aims to model the nodes in the chosen route as a coalition of players who expect to be remunerated based on their contribution. The input to this phase is a predefined route selected by a routing protocol such as AODV or DSDV [14]. The advantage of this mechanism is that it is independent of the routing method used[2]. This phase considers all selected nodes as part of a coalition and offers each node a payoff depending on the power that it saves by forwarding the packet. It is assumed that all nodes are stationary.

[2] This is because FASTER Virtual Currency Allocation will function on top of any routing protocol in the protocol hierarchy.

Consider a node $X_s$, which is the sender of a message and a node $X_d$, which is the designated destination. A relay node $X_r$ will forward the packet only if the total power used in transmission using a relay is reduced.

$$P_{sr} + P_{rd} = P_{sd} - \varepsilon \quad (3)$$

Where $P_{sr}$ is the minimum power for packet transmission from $X_s$ to $X_r$, $P_{rd}$ is the minimum power for packet transmission from $X_r$ to $X_d$, the minimum power for packet transmission from $X_s$ to $X_d$ is $P_{sd}$ and $\varepsilon$ is a positive constant, which ensures that power is always saved by the coalition.

It is the responsibility of the sending node ($X_s$) to calculate the Shapley value of each intermediate node before the packet is actually sent. For this purpose, included with the predefined route that this phase receives, is the distance between adjacent intermediate nodes.

As per the Shapley Value function, for each possible subset of the intermediate nodes (coalitions), we are required to calculate the coalition value. We use the total saved power in calculating the payoff value v(S) for each coalition S.

$$v(S) = \sum_{i \in S} 1 - \left(\frac{d_{i,i+1}}{d_{s,r}}\right)^4 \quad (4)$$

Where $d_{s,r}$ is the distance between the sender and the receiver and $d_{i,i+1}$ is the distance between adjacent hops in the path.

Once this calculation is done by the sender, it obtains the contribution of each node by substituting the coalition values in the Shapley Value function. These contributions per node are weighted by a constant amount and sent along with the selected path.

*D. Virtual Currency Allocation*

FASTER uses Virtual Currency to incentivize the mechanism and encourage participation in the system. Nodes use this currency as payment to other nodes for relaying their messages, and earn the currency when they offer their services to other nodes. This property makes Virtual Currency a necessity in the network and thus each node will aim to increase the amount of currency it has.

We assume that the user to whom the node belongs has complete control over the node, which means that a user can alter software, protocols and node equipment in order to achieve his goals. This results in 'selfish' nodes, which require motivation to cooperate.

The model that this paper proposes for charging nodes for packet forwarding service is similar to the Packet Purse Model (PPM) [15], where the sender is charged. The sender inserts into the packet the amount of currency that it must give to the nodes on the selected route for forwarding its message. In this model, the exact amount that is owed to each node is known by the Shapley Value calculation process. Due to this, each packet will contain the exact amount to be rewarded to all the nodes and will not suffer from the problem of overestimating or underestimating the quantity to be paid. In addition, each packet will also contain an indication of the amount that each node has to receive. This ensures that each node gets what it deserves for helping the coalition.

*1. Virtual Currency Representation*

The proposed Virtual Currency system can be modeled in many ways. One such method is by using virtual coins, which are objects, each of which has a unique serial number that is immune to unauthorized modification. Serial numbers contain information that each node can act upon.

The first section of the serial number represents the issuer of the credit coin and the second section identifies the explicit contract that the coin represents. The contract is a binding promise of delivery of services or products, and will contain a timestamp of date of expiry. The third section recognizes the current owner. Nobody but the owner can spend a digital coin. As a result, the ownership section can be rewritten, and the coin need not be physically sent. The fourth section of the serial number defines the value of the coin.

This method has a number of drawbacks. The overhead of simulating Virtual Currency through digital coins will be large and will significantly affect the performance of the network. In addition, this method would require established infrastructure in the form of an authority that issues coins to users, which is impractical in this case as Ad hoc networks should not depend on any existing infrastructure.

Citing these problems, we propose the use of a counter in each node to denote the amount of currency that it has. Counters do not suffer from the problems of transmission overhead and the need for existing infrastructure. However, to guarantee that nodes do not alter the contents of the counter and the correct counter value is maintained in each node, the counter must be maintained in a protected section, similar to that stated in [11], which does not allow access to the counter.

*2. Protection of Virtual Currency Integrity*

Protecting the integrity of Virtual Currency is a major issue when implementing the counter mechanism in each node. In order to avoid illegal modification of the counters, they must be kept out of the users' reach. The currency counter must be maintained in a protected section which is secured through trusted hardware, along with the private and public keys of the section, which should also be shielded from alteration by the user. In addition, the protected section must also maintain a database of information of other nodes' protected section. This includes storing of a session key [16] for each node that a node communicates with. The session key is used in the Public Key Infrastructure (PKI) cryptography mechanism to ensure that the packets containing Virtual Currency are not tampered with, and each node accepts only its share of currency.

The issue of each node taking only the amount of currency that is allotted to it by the Shapley Value calculation is also a very relevant one. To ensure that this is the case, we propose a design based on PKI. The FASTER VCA will include a header that is divided into sections, each of which will contain identifying information about a hop on a selected route and the amount of currency that is owed to it. Each section would be encrypted with the public key of the node that it represents. By PKI design, each node will be allowed access only to the section that it is concerned with. This is because the use of a node's private key on a packet to decrypt will only reveal information about currency allocation that is relevant to it. All other sections will still remain encrypted.

## IV. SIMULATION AND EVALUATION

### A. Simulation Environment

In order to observe the effects of the proposed system model on the performance of the network, we implemented FASTER in Python using the Networkx package. Our aim was to show that the Ad hoc network would be benefited when all nodes used FASTER.

In the simulation, each node had a 250m communication range. FASTER was simulated in a 500m × 500m area. The simulation was conducted with 10, 15, and 20 nodes and each of them was randomly positioned on the grid. Nodes were assumed to be stationary. All nodes had an initial energy of 100J. Transmission Power cost was 1.4W and receiving power cost was 1W. Idle Power cost was taken to be 0.83W. This energy model used was the same as that of CAP-SV. Every node had a chance of sending a packet that was decided by a coin toss with a certain probability. If a node decided to send a packet across the network, then a suitable route through different hops was selected, and FASTER was used to find a fair currency allocation.

### B. Results

Node richness is the amount of currency that a node has at a given point of time. In Figure 1, we can see that the discrepancy between different nodes in terms of richness is high. Since the amount of currency that a node gains is not proportional to its contribution in the network, some nodes may become very rich quickly due to a topographical advantage. Since power reduction is not considered, all nodes are paid equally, and these nodes become much richer than others.

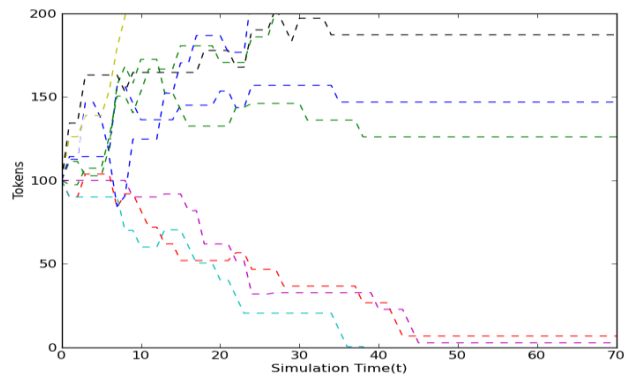

Fig. 1. Simulation Time vs. Richness without FASTER

In Figure 2, we can see that most nodes die out early when FASTER is not used, and this leads to low network performance, as most nodes with low remaining power will refuse to relay the messages of other nodes.

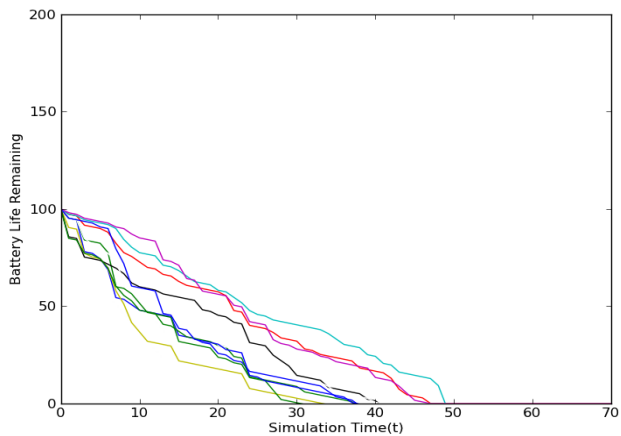

Fig. 2. Simulation Time vs. Remaining Battery Life without FASTER

When using FASTER, there is a much greater equality among nodes with respect to battery life as well as richness. This is because nodes are given encouragement to participate in the forwarding of messages by allocating them some currency. It is evident from Figure 3 that the cooperative game theory aspect of FASTER introduces a greater equality among nodes in terms of the amount of currency that they possess. Nodes which save greater power and in turn use greater power in relaying will be rewarded more currency. Nodes which are rich will not have the power to continuously send messages as they have spent more battery in becoming rich.

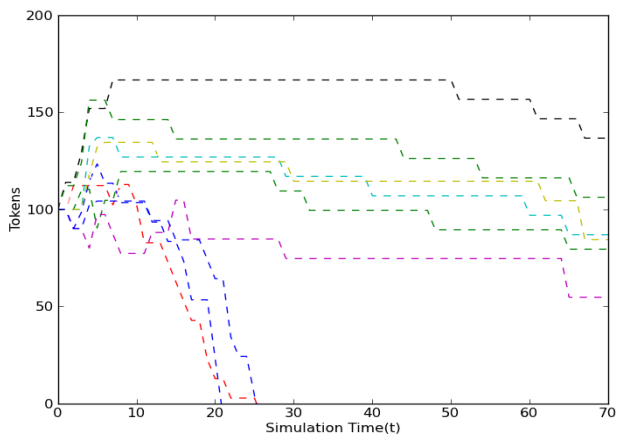

Fig. 2. Simulation Time vs. Richness with FASTER

From Figure 4, it is apparent that network performance has significantly improved due to the use of FASTER. Nodes last much longer in the network, and this supports the act of relaying information, which leads to improved network performance. With the inclusion of Shapley Value based calculation, nodes are given incentive to provide their services to the network. When all nodes exhibit this behavior, each node is able to last longer than it would have if it had been selfish.

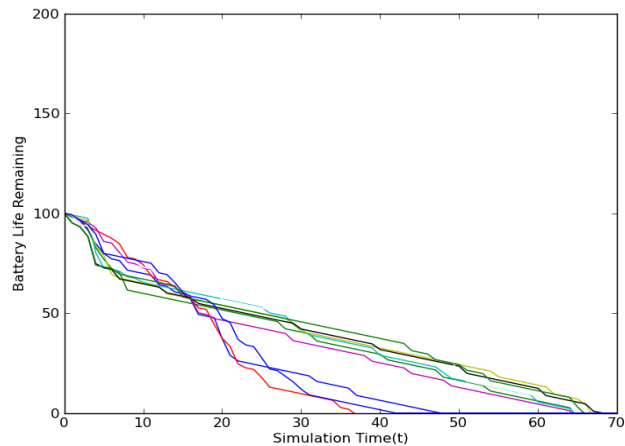

Fig. 4. Simulation Time vs. Remaining Battery Life with FASTER

## V. CONCLUSIONS AND FUTURE WORK

Lack of centralized control in Ad hoc networks implies that the behavior of the autonomous nodes that constitute the network has an enormous impact on network performance. The presence of self-governing nodes also implies that cooperation must be stimulated in the system in order to improve the service that each node receives. In this paper, we propose a protocol FASTER that provides a fair and incentivized mechanism to improve cooperation among nodes and hence improve network productivity.

We present a scheme to determine the payoff that each node has to receive in a partnership using the cooperative game theory concept of Shapley Value. This ensures that each node is remunerated based on the services that it provides to the network.

After the calculation of the Shapley Value, we also propose a method to distribute Virtual Currency effectively to the selected relay nodes in a secured manner so as to prevent any nodes from taking more currency than it deserves.

Our scheme addresses the issue of how to determine the payoff for each node, as well as the issue of redeeming the earnings from the network. We prove by simulation that Ad hoc networks will benefit from the use of FASTER.

Scalability of the protocol is an area for future work. This is due to the fact that the Shapley Value calculation involves computation based on the factorial of the selected number of hops. In a large network, this could be a problem.

REFERENCES

[1] Zhong, Sheng, Jiang Chen, and Yang Richard Yang. "Sprite: A simple, cheat-proof, credit-based system for mobile ad-hoc networks." In INFOCOM 2003. Twenty-Second Annual Joint Conference of the IEEE Computer and Communications. IEEE Societies, vol. 3, pp. 1987-1997. IEEE, 2003.

[2] Zhu, Haojin, Xiaodong Lin, Rongxing Lu, Yanfei Fan, and Xuemin Shen. "Smart: A secure multilayer credit-based incentive scheme for delay-tolerant networks. "Vehicular Technology, IEEE Transactions on 58, no. 8 (2009): 4628-4639.

[3] Chen, Bin Bin, and Mun Choon Chan. "Mobicent: A credit-based incentive system for disruption tolerant network." In INFOCOM, 2010 Proceedings IEEE, pp. 1-9. IEEE, 2010.

[4] AuYoung, Alvin, Brent Chun, Alex Snoeren, and Amin Vahdat. "Resource allocation in federated distributed computing infrastructures." In Proceedings of the 1st Workshop on Operating System and Architectural Support for the On-demand IT InfraStructure, vol. 9. 2004.

[5] S. Das, C.E. Perkins and E. M. Royer. Ad hoc On Demand Distance Vector(AODV) Routing. Mobile Ad-hoc Network (MANET) Working Group, IETF, January 2002.

[6] Yang, Yaling, Jun Wang, and Robin Kravets. "Designing routing metrics for mesh networks." In IEEE Workshop on Wireless Mesh Networks (WiMesh). 2005.

[7] Cai, Jianfeng, and Udo Pooch. "Allocate fair payoff for cooperation in wireless ad hoc networks using shapley value." In Parallel and Distributed Processing Symposium, 2004. Proceedings. 18th International, p. 219. IEEE, 2004.

[8] Gomez, Javier, Andrew T. Campbell, Mahmoud Naghshineh, and Chatschik Bisdikian. "PARO: Supporting dynamic power controlled routing in wireless ad hoc networks." Wireless Networks 9, no. 5 (2003): 443-460.

[9] Chen, Benjie, Kyle Jamieson, Hari Balakrishnan, and Robert Morris. "Span: An energy-efficient coordination algorithm for topology maintenance in ad hoc wireless networks." In Proceedings of the 7th annual international conference on Mobile computing and networking, pp. 85-96. ACM, 2001.

[10] Srinivasan, Vikram, Pavan Nuggehalli, Carla F. Chiasserini, and Ramesh R. Rao. "Energy efficiency of ad hoc wireless networks with selfish users." InEuropean Wireless Conference 2002 (EW2002). 2002.

[11] Buttyan, Levente, and Jean-Pierre Hubaux. "Nuglets: a Virtual Currency to stimulate cooperation in self-organized mobile ad hoc networks." (2001).

[12] Srinivasan, Vikram, Pavan Nuggehalli, Carla F. Chiasserini, and Ramesh R. Rao. "Cooperation in wireless ad hoc networks." In INFOCOM 2003. Twenty-Second Annual Joint Conference of the IEEE Computer and Communications. IEEE Societies, vol. 2, pp. 808-817. IEEE, 2003.

[13] David Kotz, Calvin Newport, Robert S. Gray, Jason Liu, Yougu Yuan, and Chip Elliott. "Experimental evaluation of wireless simulation assumptions", Dartmouth Computer Science Technical Report TR2004-507 June 2004.

[14] Perkins, Charles E., and Pravin Bhagwat. "Highly dynamic destination-sequenced distance-vector routing (DSDV) for mobile computers." ACM SIGCOMM Computer Communication Review 24, no. 4 (1994): 234-244.

[15] Buttyán, Levente, and Jean-Pierre Hubaux. "Enforcing service availability in mobile ad-hoc WANs." In Proceedings of the 1st ACM international symposium on Mobile ad hoc networking & computing, pp. 87-96. IEEE Press, 2000.

[16] Perrig, Adrian, John Stankovic, and David Wagner. "Security in wireless sensor networks." Communications of the ACM 47, no. 6 (2004): 53-57.